# PUZZLES OF THE INTERPLANETARY MAGNETIC FIELD IN THE INNER HELIOSPHERE

**Olga Khabarova and Vladimir Obridko**

Heliophysical Laboratory, Institute of Terrestrial Magnetism, Ionosphere and Radiowave Propagation RAS (IZMIRAN),
Troitsk, Moscow 142190, Russia; habarova@izmiran.ru

ABSTRACT

Deviations of the interplanetary magnetic field (IMF) from Parker's model are frequently observed in the heliosphere at different distances $r$ from the Sun. Usually, it is supposed that the IMF behavior corresponds to Parker's model overall, but there is some turbulent component that impacts and disrupts the full picture of the IMF spatial and temporal distribution. However, the analysis of multi-spacecraft in-ecliptic IMF measurements from 0.29 AU to 5 AU shows that the IMF radial evolution is rather far from expected. The radial IMF component decreases with the adiabatic power index ($|B_r| \propto r^{-5/3}$), the tangential component $|B_t| \propto r^{-1}$, and the IMF strength $B \propto r^{-1.4}$. This means that the IMF is not completely frozen in the solar wind. It is possible that turbulent processes in the inner heliosphere significantly influence the IMF expansion. This is confirmed by the analysis of the $B_r$ distribution's radial evolution. $B_r$ has a well-known bimodal histogram only at 0.7–2.0 AU. The bimodality effect gradually disappears from 1 AU to 4 AU, and $B_r$ becomes quasi-normally distributed at 3–4 AU (which is a sign of rapid vanishing of the stable sector structure with heliocentric distance). We consider a quasi-continuous magnetic reconnection, occurring both at the heliospheric current sheet and at local current sheets inside the IMF sectors, to be a key process responsible for the solar wind turbulization with heliocentric distance as well as for the breakdown of the "frozen-in IMF" law.

## 1. INTRODUCTION

Since Parker's model of the solar wind plasma flow had become commonly accepted (Parker 1958, 1963), the general structure of the interplanetary magnetic field (IMF) was supposed to be well known. Any solar wind model refers to Parker's model (see a review by Echim et al. 2011 and references therein), and the spiral form of the IMF suggested by Parker is a "corner-stone" of solar physics.

During the last several decades, it has been convincingly shown that Parker's model in its canonical view (when a steady and purely radial outflow of the coronal plasma with a "frozen-in" magnetic field is expected) as well as other radial expansion models using the Parker solution are acceptable for both the solar wind speed behavior forecast and for the IMF sign (polarity) prediction (Obridko et al. 1996; Zhao et al. 2006; Belov et al. 2006). Permanent improvement of the forecasting technique allows us to reach a success rate of prediction ~0.6–0.9 (Arge & Pizzo 2000; Luo et al. 2008).

Meanwhile, some unexpected features of the solar wind speed radial evolution has been experimentally found (McGregor et al. 2011). A simple comparison of theoretical and empirical models shows that even the most easily predictable parameter, the solar wind speed at 1 AU, is forecasted better by empirical models rather than by models using Parker's solution (Owens et al. 2008). This fact may be taken into account when we consider reasons for the failure of the IMF behavior predictions based on radial outflow models. Those failures happen so often and regularly that they are worthy to be considered as a separate problem, demanding extended investigations.

Indeed, although the sign of the IMF at 1 AU can be predicted rather successfully (see, for example, http://www.swpc.noaa.gov/ws/ NOAA Space weather prediction center page), the IMF strength forecasts using the solar wind radial outflow models face significant problems (see, for example, Riley 2007). The growing volume of experimental data obtained by various missions (such as *Ulysses*, *STEREO*, etc.) manifest dissimilarity between the observed IMF characteristics and the calculated ones (Schwadron 2002; Belov et al. 2006; Riley & Gosling 2007; Sternal et al. 2011; Ruiz et al. 2011). Heliospheric magnetic topology probes using Jovian electrons, performed as an independent test, confirm impossibility to explain the experimentally obtained results by classical IMF expansion along the Parker spiral (McKibben et al. 2005; Owens et al. 2010).

The reports listed above show that mismatches between the expected IMF characteristics and observations cannot be simply determined by purely statistical reasons, i.e., by Gaussian distribution of the IMF values (or directions) with some abmodality. There are at least two problems: a problem of the IMF direction deviations from the Parker spiral and a problem of rather poor correspondence of the IMF strength simulation results to measurements in space.

Usually, the IMF deviations from the expected direction are considered to be variations mainly of the Parker spiral-directed magnetic field due to the influence of coronal mass ejections (CMEs) and corotating interaction regions (CIRs) on the heliospheric topology or as a result of turbulent processes (Ragot 2006; McGregor et al. 2008; Borovsky 2010; Owens et al. 2010).

One of the most probable ways for this problem solving was suggested by Schwadron & McComas (2005), who supposed an existence of the sub-Parker spiral IMF determined by the motion of open magnetic field foot points at the Sun across the coronal hole boundary. So, it is possible to find quite reasonable physical explanations for the discussed phenomenon and to apply needed corrections to IMF models.

The problem of "the IMF strength" cannot be solved so easily. If one takes synoptic maps and evaluates the line-of-sight magnetic field at the source surface in a radial IMF component $B_r$ at 1 AU according to Parker's model (using the $r^{-2}$ IMF strength decrease law, where $r$ is the distance from the Sun), then the calculated field strength is found to coincide poorly with the observed $B_r$ (or $B_x$ in the Geocentric Solar Ecliptic System (GSE)), being much lower (Obridko et al. 1996). The application of some correction factors (both for the solar magnetograms and for the IMF extension law in the solar wind) and the use of hybrid empirical–theoretical models can improve the IMF strength forecasts (Wang & Sheeley 1995; Zhao & Hoeksema 1995; Obridko et al. 2004). Meanwhile, all of the "technical" solutions, such as modifying the computation scheme or shifting the source surface height (Lee et al. 2011), do not throw light on the physical nature of the phenomenon and cannot even provide calculated $B_r$ values comparable with measured ones at an acceptable statistical level, as was convincingly shown by Riley (2007).

An increasing number of reports about rather poor correspondence of the radial expansion models (traditionally used for the calculation of the IMF characteristics in the inner heliosphere) to observations have stimulated us to compare the observed IMF characteristics at different AU with simulations. We mainly consider here the problem of "the IMF strength discrepancy" in different aspects.

For the best understanding of the picture of the IMF temporal and spatial distribution in the inner heliosphere, we used OMNI2 daily data as well as data from different spacecraft (*Helios 2*, *Pioneer Venus Orbiter*, *IMP8*, *Ulysses*, and *Voyager 1*) for the period from 1976 to 2009. Certainly, no one spacecraft provides data for the whole time range, but the number of measurements used is enough to make statistically proved conclusions.

## 2. EVOLUTION OF THE INTERPLANETARY MAGNETIC FIELD STRENGTH WITH HELIOCENTRIC DISTANCE: LOOKING FOR A POWER LAW

Most successful solar wind models use a combined polytropic index, calculated based on spacecraft measurements at some AU (see Usmanov et al. 2011 and references therein). This fact may indicate that a key problem of the IMF modeling is not in deviations from the Parker spiral, but instead in the use of an incorrect expansion law. In other words, our understanding of the full IMF picture in the inner heliosphere is insufficient, and, possibly, we have to consider different approaches for the description of the solar wind plasma and magnetic field extension from the Sun.

The first systematic attempt to reveal the real $B(r)$ law was made by Totten et al. (1995). They considered the polytropic equation of the ideal gas:

$$\frac{p}{\rho^\alpha} = const., \quad (1)$$

where $p$ is a sum of plasma pressure and magnetic pressure, $\rho = nm_{proton}$ is the mass density, and $\alpha$ is a polytropic index, which may be derived from the equation

$$\frac{\alpha\beta - 2\lambda}{\delta + \beta - \alpha\beta} = \beta_p. \quad (2)$$

Here, $\beta_p$ is a plasma beta (the plasma pressure to magnetic pressure ratio), $\beta$ is a power index for number density $n(r)$, $\delta$ is a power index for the solar wind temperature $T(r)$, and $\lambda$ is a power index of the IMF strength $B(r)$. As one can see, $\alpha$ is a function of $\lambda$:

$$\alpha = \frac{\beta_p(\delta + \beta)}{\beta(1 + \beta_p)} + \frac{2}{\beta(1 + \beta_p)}\lambda. \quad (3)$$

**Table 1**
Details of the IMF Measurements and Number of Cases (hours) Used for Calculating $|B_r|$, $|B_t|$, and $B$ in Figures 1 and 2

| $r$ (AU) | Spacecraft, Year | No. of Cases |
|---|---|---|
| 0.29–0.4 | *Helios 2*, 1976 | 2557 |
| 0.4–0.6 | *Helios 2*, 1976–1978 | 2506 |
| 0.7 | *Pioneer Venus Orbiter*, 1978–1979 | 6942 |
| 1 | *IMP8*, 1976–1979 | 17459 |
| 1–2 | *Voyager 1*, 1977 | 2444 |
| 2–3 | *Voyager 1*, 1978 | 1916 |
| 3–4 | *Voyager 1*, 1978 | 1820 |
| 4–5 | *Voyager 1*, 1978–1979 | 2521 |

Using the $\beta$, $\delta$, and $\beta_p$ values obtained by Totten et al. (1995) from the *Helios 1* spacecraft data, one can find that the second term in Equation (3) is about twice greater than the first one, so $\alpha$ is highly determined by the magnetic field power index $\lambda$. This demonstrates the importance of experimentally determining $\lambda$ based on the full database for the IMF at different AU.

Totten et al. (1995) found the polytropic index value to be $\alpha = -1.58 \pm 0.06$. This is less than the adiabatic value of 5/3, but greater than the isothermal value of 1. As shown in Totten et al. (1995), $B \propto r^{-1.64 \pm 0.09}$, and there is no significant dependence of $\lambda$ (and even $\alpha$) on the solar wind speed, which contradicts the solar wind behavior predicted by Parker's model.

Parker's theory states $B_r \times r^2$ conservation along a radial direction, and this invariant (based on the $r^{-2}$ law of the IMF strength decrease) is considered in space physics very often, for example, for $B_r$ recalculation from one heliocentric distance to another (Balogh et al. 1995; Smith et al. 2000). In that case, it would be reasonable to derive based on multi-spacecraft measurements not only the $\lambda$ index for $B(r)$, but also the radial and tangential IMF components' contribution to the power law.

Further comparison of the observed $B_r$ IMF magnitudes at different heliocentric distances with the $B_r$ values predicted by Parker's model would be useful for understanding the cause of the failures of the radial outflow models.

To find how the IMF changes with heliocentric distance, we have tried to select several spacecraft, covering a possibly long distance simultaneously. Of course, the latter was not fully achievable, but we found that the *Helios 2*, *Pioneer Venus Orbiter*, *IMP8*, and *Voyager* spacecraft maximally intersected during 1976–1979, when they measured the IMF characteristics from 0.29 AU to 5 AU. For some spacecraft, it was impossible to obtain data with accuracy better than one hour, so we used hourly IMF data (the radial component $B$ as well as absolute values of signed radial $B_r$ and tangential $B_t$ component), averaged for our purposes. The *IMP8* and the *Pioneer Venus Orbiter* spacecraft data were averaged over time. For spacecraft with elongated orbits, the hourly IMF components' values were averaged not only temporally, but also spatially (see details in Table 1).

Statistical analysis shows that such an approach is suitable for our aims. For example, according to the *Pioneer Venus Orbiter* spacecraft measurements in 1978–1979, the module $B_r$ mean value $<|B_r|> = 6.38 \pm 3.61$ nT at 0.7 AU practically coincides with the same parameter, based on the other spacecraft (*Helios 2*) hourly data, averaged over time (1978–1979) and space (from 0.6 AU to 0.8 AU): $<|B_r|> = 6.32 \pm 4.07$ nT.

Results of the study of $|B_r|$, $|B_t|$, and $B$ versus $r$ are shown in Figure 1, where different spacecraft are marked with different labels (Figures 1(a)–(c)). The corresponding trend lines can be found in the right panel of Figure 1 (see Figures 1(d)–(f)). We use here the least-squares fitting method, taking the logarithms of the input data, so that the results are plotted in Figures 1(d)–(f) on a logarithmic scale.

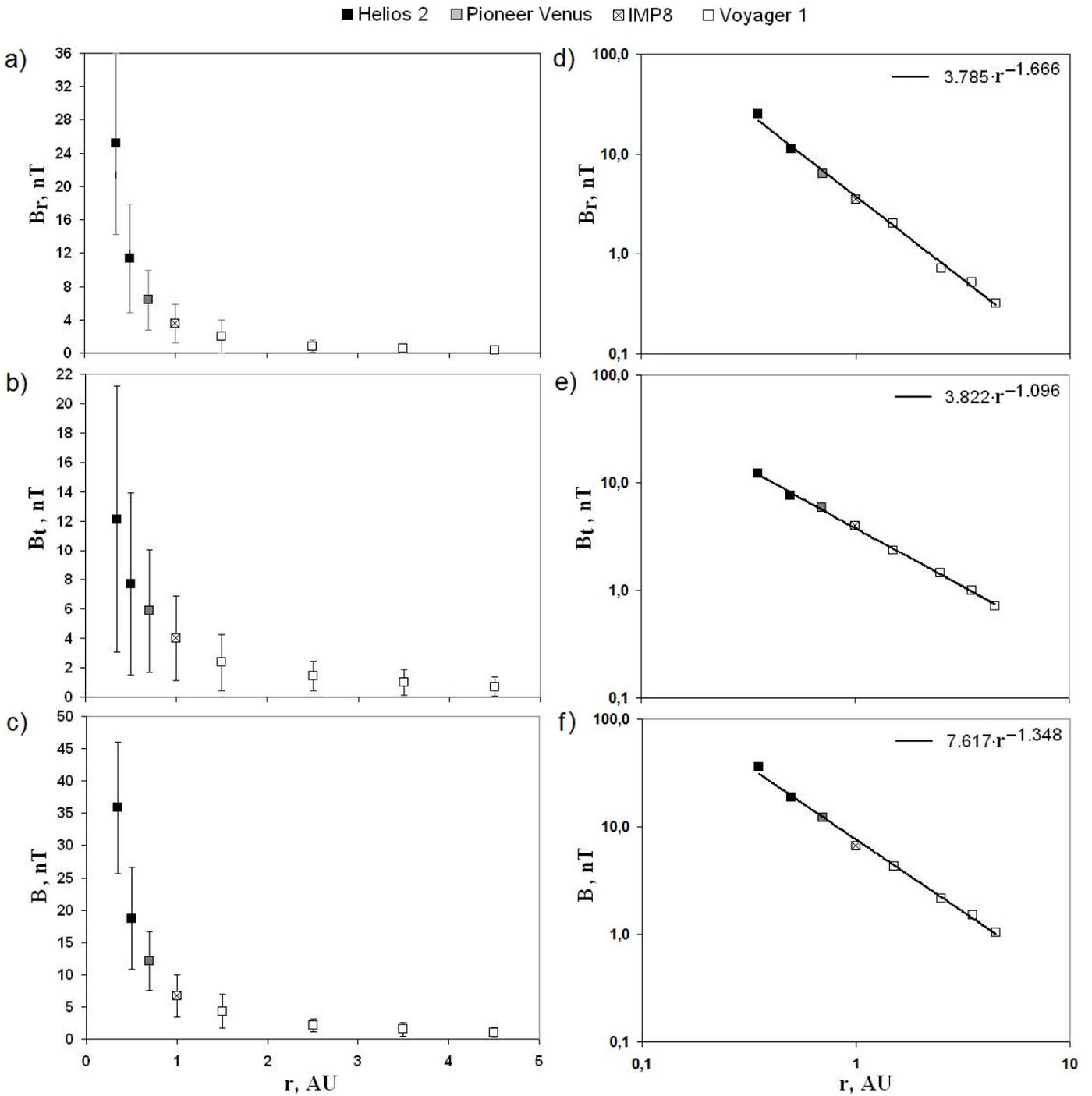

**Figure 1.** Radial variation of the strength of the IMF as observed by different spacecraft (hourly data, averaged according to Table 1). (a) Module of the radial IMF component $B_r$; (b) module of the tangential IMF component $B_t$; (c) the IMF strength $B$ at different AU; (d–f) fitting of the data shown on the left panel (a–c), a log–log scale.

Figure 1(d) represents the approximation

$$|B_r(r)| = 3.785 \times r^{-1.666} = 3.785 \times r^{-5/3}, \quad (4)$$

with a coefficient of determination $R^2 = 0.996$ (Figure 1(a)). The $R^2$ of 1.0 indicates that the regression line completely fits the data, so Equation (4) perfectly describes the change of the $|B_r|$ mean values with distance from the Sun. Analogous to Figure 1(d), one can find power laws for $|B_t|$ and $B$:

$$|B_t(r)| = 3.822 \times r^{-1.096}, \quad (5)$$

$$B(r) = 7.617 \times r^{-1.348} \quad (6)$$

(see Figures 1(e) and (f)).

The coefficient of determination is the same in all cases: $R^2 = 0.996$. This is a sign of non-randomness of the found dependences (Equations (4)–(6)).

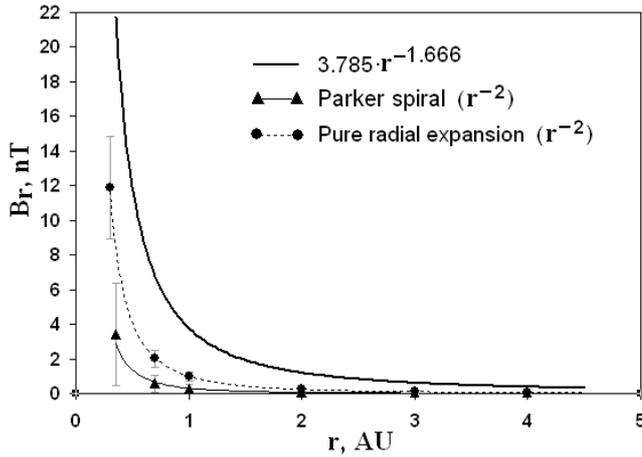

**Figure 2.** Comparison of the observed |$B_r$| variation with heliocentric distance and predictions of the radial outflow models at different AU. The |$B_r$| trend line (see Figure 1(d)) is plotted in a linear scale (the black thick curve) and compared with the $B_r$ module, calculated based on the solar magnetic field $B_R$ data, according to a simple radial expansion model (the dashed black curve, black circles) and a model taking into account the solar wind propagation time along the Parker spiral (the thin black curve, triangles).

Therefore, our ultimate aim is to explain the discovered law of IMF expansion in space from a physical point of view.

It is remarkable that the obtained power index for |$B_r$| is rather far from the expected value of −2, but |$B_t$| decreases with heliocentric distance practically as $r^{-1}$, which corresponds to the common opinion on the tangential IMF component's behavior in the heliosphere (Burlaga et al. 1998, Jian et al. 2008).

Additionally, we calculated |$B_r$| at different AU based on the measured solar $B_R$ for corresponding time periods according to the $r^{-2}$ law of the $B_r$ decrease during the radial solar wind propagation. Recalculations were made as follows: 1μT of $B_R$ corresponds to the

$$\left(\frac{r_{source}}{r \cdot r_{1AU}}\right)^2 \cdot 10^3 = \left(\frac{2.5 R_S}{r \cdot 213.6 R_S}\right)^2 \cdot 10^3 = 0.137 r^{-2} \text{ (nT)} \quad (7)$$

of $B_r$ at some distance $r$ from the Sun. $r_{source}$ (the source surface height) is taken at 2.5 solar radii ($R_S$), and 1 AU distance ($r_{1\,AU}$) is equal to $213.6 R_S$.

The correlation between the IMF sign at the Sun and at 1 AU is known to be maximal at the lag of 4.5 days (Ness & Wilcox 1966). It is commonly accepted that such a lag is a result of the solar wind propagation along the Parker spiral. So, a time delay for the quiet solar wind propagation from the Sun to some $r$ was taken for |$B_r$| calculations according to Table 2.

The simple case of the pure IMF radial expansion (Equation (7)), when the rotation effect is not taken into account, is shown in Figure 2 with circles on the dashed curve, and the thin black curve with triangles represents the result of |$B_r$| calculation at some AU along the Parker spiral as shown in Table 2. The approximation curve (Equation (4)) is also given in Figure 2 for comparison. At small distances from the Sun, the calculated IMF strengths are several times less than those measured by spacecraft.

**Table 2**
Time Lag Used for |$B_r$| Calculations in Figure 2

| $r$ (AU) | Time Delay (days) |
|---|---|
| 0.29 | 1.3 |
| 0.70 | 3.2 |
| 1.0 | 4.5 |
| 2.0 | 9.0 |
| 3.0 | 13.5 |
| 4.0 | 18.0 |

The difference is so pronounced that solar cycle or velocity effects cannot explain it. CME impact cannot be fully responsible for that either. Riley (2007) pointed out the poor correspondence between the observed and calculated IMF at 1 AU even under application of correction factors. His comparison revealed the model's failure at solar maxima. The errors also seem to be significant even in solar minima, when CME activity is low (see Figure 1 from the paper by Riley, 2007). If one performs the same comparison for other heliocentric distances, the observed difference should be more apparent closer to the Sun, according to Figure 2.

## 3. EVOLUTION OF THE DISTRIBUTION'S SHAPE OF THE RADIAL IMF COMPONENT WITH HELIOCENTRIC DISTANCE

One of the indications of uncertainty about the rules of photospheric field expansion is that the predictive models give an increased number of near-zero values of the in-ecliptic IMF strength in comparison with the number of actual observed IMF zeros at 1 AU (Belov et al 2006). One example is given in Figure 3. $B_L$ is the in-ecliptic IMF component at 1 AU. The black line with daggers is a result of the $B_L$ derivation based on a steady, radially directed flow model, using the line-of-sight magnetic field $B_R$ data. The black points are the observed values of $B_L$ at 1 AU. One can see that the number of daggers essentially exceeds the number of black points in the selected area around zero strength.

This effect is obviously a consequence of the use of a Gaussian-like $B_R$ distribution for $B_L$ calculations.

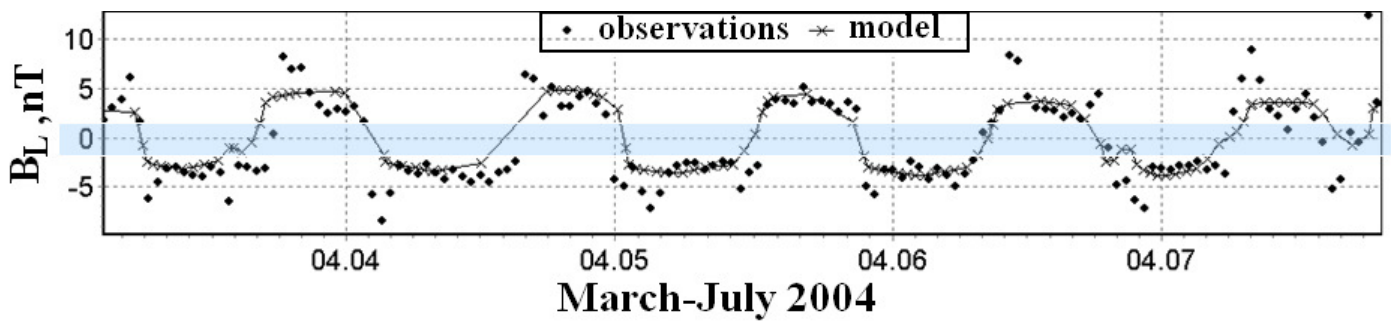

**Figure 3.** Comparison of the simulated and observed in-ecliptic IMF component $B_L$ (according to Belov et al. 2006).

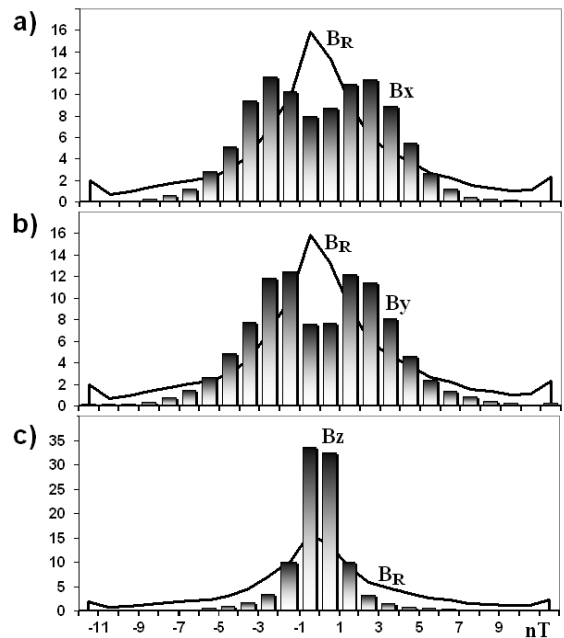

**Figure 4.** Histograms of the occurrence (in the percentage of the total) of $B_x$, $B_y$, and $B_z$ IMF components in the ecliptic plane at 1 AU according to OMNI2 daily data (bars) in comparison with the distribution of the magnetic field $B_R$ at the source surface chosen to be at 2.5 solar radii (the black line behind the bars), 1977–2009.

Indeed, at the source surface, the line-of-sight magnetic field $B_R$ is normally distributed around zero as shown in Figure 3. Meanwhile, the observed in-ecliptic IMF component at 1 AU represented by the $B_x$ and $B_y$ IMF components in the GSE coordinate system (or by the radial IMF component $B_r$, RTN) has a well-known zero-centered bimodal distribution (see Figures 4(a) and (b)). There are two peaks of distribution: one negative and one positive, while the vertical $B_z$ IMF component is normally distributed around a value of zero (Figure 4(c)).

Therefore, if we take the normally distributed line-of-sight magnetic field at the source surface $B_R$ and derive $B_r$ or in-ecliptic $B_L$ at 1 AU, then the obtained IMF component is Gaussian-like, but not bimodal, because it has the same distribution shape as the initial field $B_R$.

The effect of a bimodal in-ecliptic IMF distribution has been well known since the beginning of the space era (see, for instance, a review by Veselovsky et al. 2010). For many years, the bimodality of the $B_L$ and $B_r$ distributions has been simply explained by the existence of a clear IMF sector structure at 1 AU (Russell 2001). Objectively, the heliospheric current sheet (HCS) is rather thin in comparison with the size of the positive or negative sectors. Therefore, one expects to observe many positive and negative in-ecliptic IMF strength values (belonging to the sectors) compared to a small number of the IMF strength zeros, associated with the HCS crossing. It is assumed that the $B_r$ distribution remains bimodal at any heliocentric distance. As a result, spatial changes of the distribution view have not been analyzed and discussed in scientific literature.

Belov et al. (2006) noted that the IMF in-ecliptic distribution at 1 AU looks absolutely different from the $B_R$ histogram.

If we trust the most popular MHD coronal expansion models beginning with Biermann (1957), Parker (1958), and Pneumann & Kopp (1971) and including current models (Schwadron & McComas 2005), then the distribution of the magnetic field in the photosphere or at the source surface along the ecliptic plane of the Sun (line-of-site magnetic field) should have the same view as the in-ecliptic IMF at 1 AU. We expect that the projection of the Earth onto the Sun crosses the warped heliomagnetic equator approximately as many times as the HCS crosses the Earth orbit at 1 AU. Consequently, we should observe one-type distributions of the corresponding magnetic fields, which do not change with heliocentric distance.

Meanwhile, statistical analysis reveals a significant change in the distribution of the $B_r$ view with distance from the Sun. Figure 5 demonstrates the transformation of the $B_r$ distribution with heliocentric distance compared to the $B_r$ distribution, calculated for the same period based on a radial outflow model, using the line-of-sight $B_R$ component values from Wilcox Solar Observatory synoptic charts. The model uses the $r^{-2}$ law and takes into account the IMF helicity as shown in the previous section.

We have chosen this model, untarnished by any corrections, just to show the inability of the radial outflow models to produce a bimodal $B_r$ distribution and correct $B_r$ strengths. Any such radial outflow model will give good results in the IMF sign prediction at 1 AU, but will poorly coincide with the $B_r$ strength, observed, for example, by Pioneer Venus Orbiter at 0.7 AU from 1978 December 5 to 1992 December 31 (Figure 5(a)) and by the OMNI2 database-formative spacecraft IMP8 and ISEE 3 at 1 AU for the same period (Figure 5(b)).

The derived $B_r$ values are obviously lower in both cases and are quasi-normally distributed. The bimodality of the $B_r$ distribution is highly apparent both at 0.7 AU and 1 AU, but at the Earth orbit, the IMF is weaker and the two humps are closer to each other than at the Venus orbit.

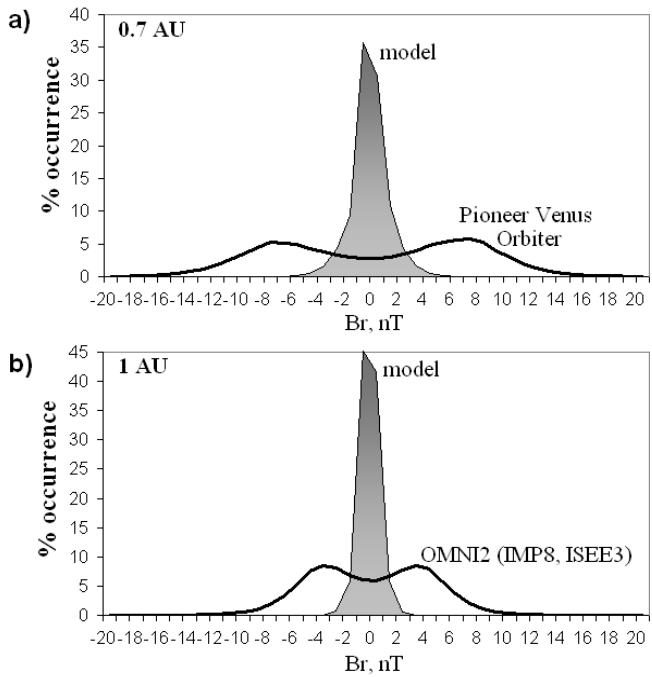

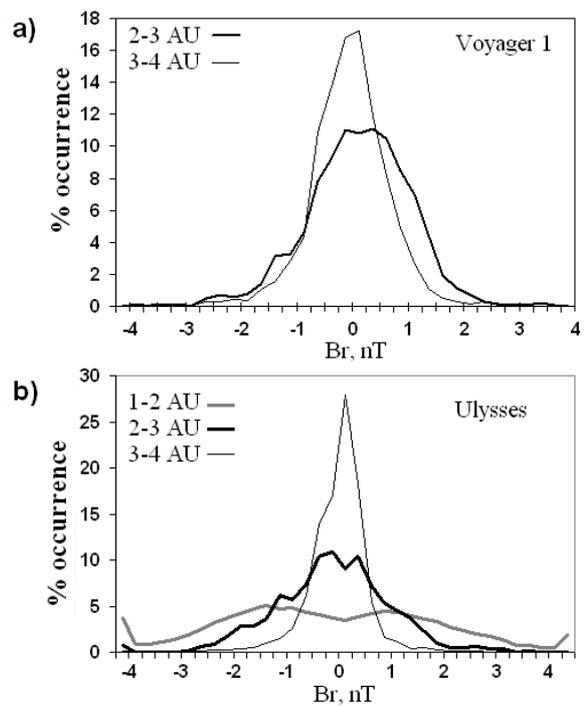

**Figure 5.** View of the $B_r$ histogram at 0.7 AU and 1 AU. (a) Black line is the $B_r$ distribution at 0.7 AU (*Pioneer Venus* hourly data, 1978–1992). The filled gray curve represents $B_r$, calculated based on the solar field $B_R$ for the same period, according to Parker's model. (b) The same as in (a), but for 1 AU (OMNI2 daily data, 1978–1992).

The continued shift of the two humps of the $B_r$ distribution towards zero with increasing distance from the Sun is shown in Figure 6. The *Voyager 1* spacecraft measured the IMF strength close to the ecliptic plane for the distances analyzed here. The $B_r$ histogram at 2.00–2.99 AU (from 12:00 1978 January 8 to 11:00 1978 April 24) still demonstrates a weak bimodality effect, but the distribution at 3.00–3.99 AU (from 12:00 1978 April 24 to 11:00 1978 August 24) shows no sign of this (see Figure 6(a)).

To check this result, we extracted periods from all available *Ulysses* spacecraft data for the near-ecliptic plane latitudes (lower than ±40°) and drew $B_r$ histograms for the distances from 1 AU to 4 AU in Figure 6(b). One can see that the effect of the $B_r$ bimodality is eliminated when the heliocentric distance increases, as in Figure 6(a). Each time the $B_r$ bimodality vanished, we looked for the effect, considering the smallest histogram bin size possible. The analysis confirms that the two-humped $B_r$ distribution indeed became Gaussian at 3–4 AU as was measured by both *Voyager 1* and *Ulysses*.

To summarize, we conclude that the radial IMF component in the ecliptic plane seems to be normally distributed at the Sun, but then it becomes bimodally distributed somewhere between the Sun and the Earth, after which its bimodality gradually vanishes with distance. From this point of view, it would be interesting to investigate the view of the $B_r$ distribution closer to the Sun.

Such a possibility was suggested by the IMF measurements obtained from the *Helios 2* mission. *Helios 2* had a very elongated orbit; it was a unique space probe that approached the Sun closer than Mercury, at a distance of 0.29 AU. We tested $B_r$ at four distance intervals (as shown in Figure 7) for all of the available hourly data between 1976 January 16 and 1980 March 5.

**Figure 6.** Dependence of the shape of the radial IMF component histogram on heliocentric distance. (a) The black line is the $B_r$ distribution at 2–3 AU, and the gray line represents the $B_r$ distribution at 3–4 AU (*Voyager 1* hourly data, 1978). (b) The gray line is the $B_r$ distribution at 1–2 AU, the black thick line is the same at 2–3 AU. The black thin line represents the $B_r$ distribution at 3–4 AU. The distributions are calculated for low heliolatitudes (lower than ±40°) according to *Ulysses* hourly data for 1990–2008.

The $B_r$ distribution has a deep hole around zero from 0.29 AU to 0.4 AU (Figure 7(a)). The depth of the hole depends on the distance from the Sun. At distances 0.8–1.0 AU (Figure 7(b)), the $B_r$ is approximately normally distributed compared with the distribution at 0.29–0.4 AU in Figure 7(a). This confirms the existence and non-randomness of the discovered effect of a gradual transformation of the bimodal $B_r$ distribution to nearly normal with heliocentric distance increase. The effect has not been investigated before and, obviously, this is a manifestation of the complexity of the IMF picture in the inner heliosphere.

## 4. DISCUSSION

As it follows from the obtained results, global physical properties of the IMF in the inner heliosphere are still not fully investigated. Let us start with an explanation of the unexpected dependence of the IMF radial component bimodality effect on heliocentric distance. Statistically, it is easy to find that the $B_r$ distribution at 1 AU, and even at the Venus orbit (0.7 AU) shown in Figure 5, is a case of the mixture of two normal distributions with the same variance but different means. This agrees with the explanation of the bimodality effect by the existence of sunward/outward IMF sectors with approximately equal impacts on the $B_r$ distribution. Physically, this means that an apparent sector structure and a very clear occurrence of the IMF neutral line

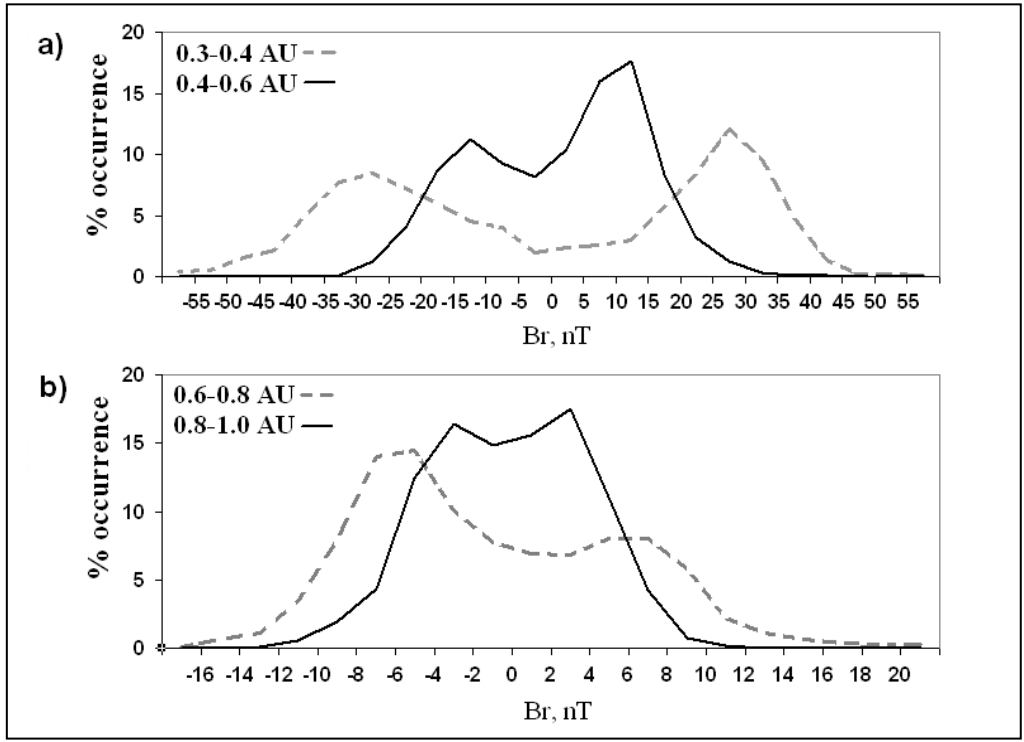

**Figure 7.** $B_r$ distribution as measured by the *Helios 2* spacecraft from 0.29 AU to 1 AU (hourly data, 1976–1980).

(the heliospheric current sheet) exist at corresponding heliocentric distances.

One can see from Figures 6 and 7 that the unexpected IMF picture is observed closer to the Sun or, on the contrary, farther from the "gold distance", 0.7–1.0 AU. Confirmation of this point is seen in Figure 7(a), where the $B_r$ distributions at 0.3–0.4 AU (dashed gray curve) consist of two unimodal distributions with a plateau between them. This plateau is not caused by the flanks of the unimodal distributions and cannot be represented by their simple superposition. This means that there is a rather stable occurrence of the IMF neutral lines at such distances from the Sun, i.e., it is a sign of the significant contribution of weak magnetic fields to the $B_r$ distribution. Averaged daily, the same distribution becomes purely three modal.

Simultaneously, the observed rapid disappearance of the Br bimodality at distances slightly farther than the "gold distance" (as seen in Figure 6) may demonstrate an increase in the complexity of the solar wind with distance from the Sun. Statistically, it resembles an approach to the normal distribution of random variables clustered around zero.

Twisting of the Parker spiral at the considered heliocentric distances is not sufficient to cause this effect, as, according to Parker's theory, the IMF lines become perpendicular to the radial direction far from 5 AU (approximately at 15 AU), but the $B_r$ distribution becomes quasi-normal at 3–4 AU. Moreover, as experimental observations show (see the introduction), the IMF geometry in the inner heliosphere is more radial than expected; hence, we have to look for another explanation for the observed phenomenon.

Several authors simultaneously came to the conclusion that the solar wind becomes more turbulent, and the sector structure becomes more and more complex with increasing heliocentric distance (Roberts et al. 2005; Burlaga et al. 2003).

We confirm this statement and add that the IMF structure seen in the ecliptic plane at 1 AU is rather unique. Obviously, any dependencies found at 1 AU should not be extended to the inner heliosphere overall, as the IMF large-scale picture significantly changes at very short distances.

Our results demonstrate that the radially extending solar wind approach fails for the IMF versus $r$ computations in the inner heliosphere, even with regards to the IMF helicity (see Figures 1 and 2). The obtained slope of the $|B_r(r)|$ approximation curve equals the adiabatic value of $-5/3$ (instead of the expected value of $-2$), but $|B_t|$ decreases as previously predicted: $|B_t| \propto r^{-1}$. So, it is easy to explain the obtained power index for $B$ ($\lambda = 1.35$) by the unexpected contribution of the $B_r$ component in $B$. It is remarkable that the approximation formulas for $B_r$ and $B_t$ have the same coefficient of $3.8 \pm 0.02$ (see Equations (4) and (5)), so the contributions of the radial and azimuthal IMF components to the IMF module are equal, which confirms the adequacy of the found expansion law (Equation (6)). This allows us to estimate the solar wind polytropic index $\alpha$ according to Equation (3) as $\alpha = 0.49 + 0.89 = 1.38$. As mentioned above, the $\alpha$ value is very close to $\lambda$.

The difference between the $|B_r|$ calculated based on Parker's model and the observed $|B_r|$ is very significant at small distances from the Sun. Meanwhile, the $|B_r(r)|$ and $B(r)$ are satisfactorily consistent with Parker's model at distances from 5 AU, as was previously shown by Burlaga et al. (1998, 2002).

As follows from the key figure in the paper by McKibben et al. (2005; see also Figure 3.2 in the *Ulysses* mission proposal http://ulysses.jpl.nasa.gov/2005-Proposal/UlsProp05.pdf), the magnetic field deviations from the Parker spiral required for a successful magnetic connection to produce observed electron jets at *Ulysses* depend on the heliocentric distance: the closer to the Sun, the higher those deviations.

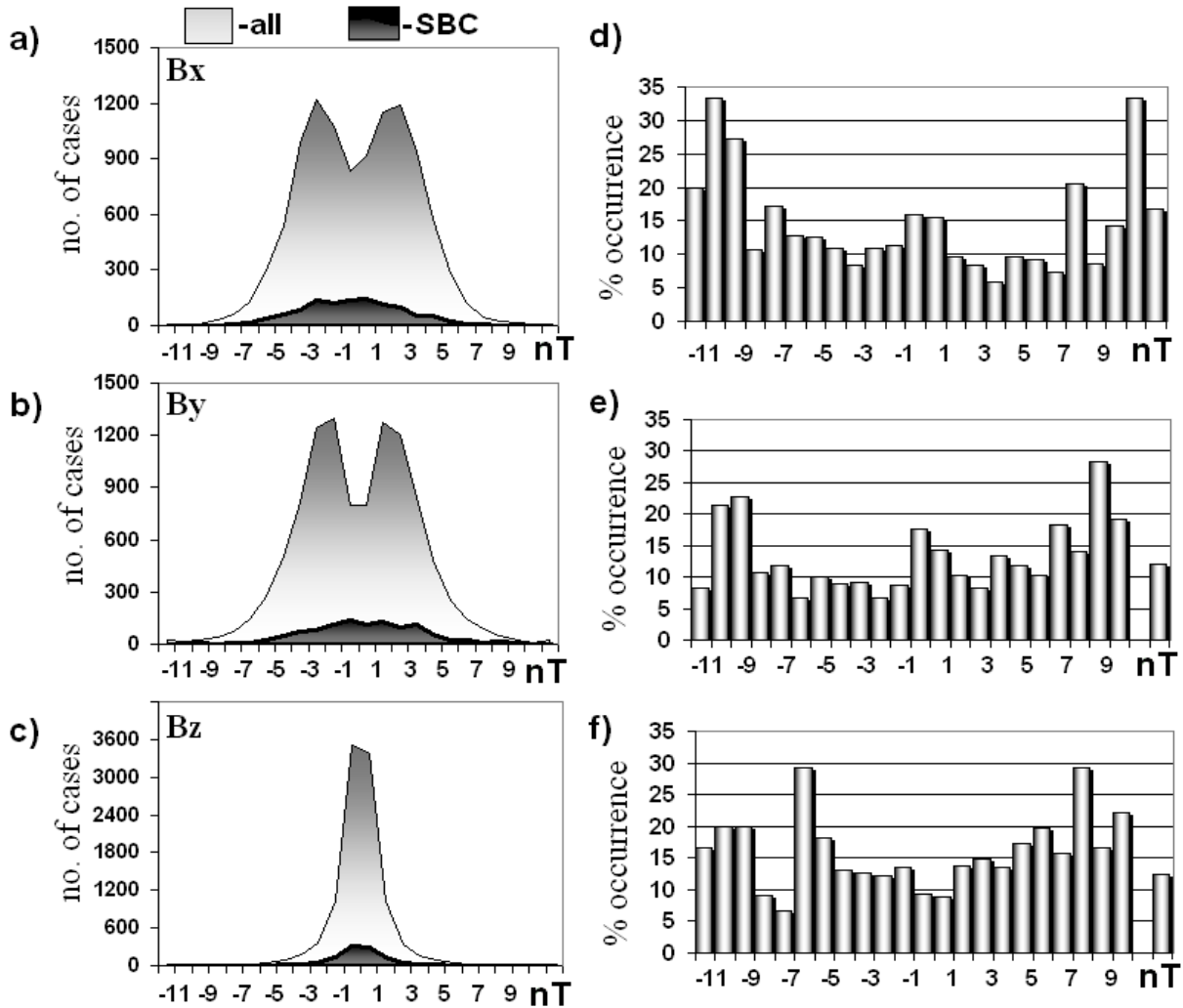

**Figure 8.** Sector boundaries' input into the interplanetary magnetic field pattern. (a–c) Histograms of the occurrence of the three IMF components ($B_x$, $B_y$, $B_z$ — GSE) at 1 AU (gray filled curves) in comparison with the histograms of the same parameters for sector boundary crossing days (black filled curves). (d–f) Percentage input of the sector boundaries into each span of the corresponding IMF histograms (a–c). OMNI2 daily data for 1977–2009.

This fact may be considered as an additional confirmation of the obtained results, which indicate some kind of independence of the IMF expansion in space from the solar wind plasma extension. Possibly, it is a sign of some break of the "frozen-in IMF" law. In other words, the IMF lines are not fully imbedded in the solar wind plasma, but there is also a component, weakly dependent on the plasma behavior. That is the reason why the solar wind speed in the inner heliosphere obeys the Parker's theory satisfactorily, but the IMF does not.

Conditions when the IMF is not frozen in the solar wind are most expressed at the HCS. This leads to a magnetic reconnection which may be repeating and prolonged, as was experimentally shown by Phan et al. (2006, 2009), and confirmed by Zharkova & Khabarova (2012) through the comparison of simulations and observations. Magnetic reconnection at current sheets changes field structure in heliosphere. Gosling et al. (2007) mentioned this effect in their work. They believe that the number of magnetic lines originally connected with the Sun is reduced by prolonged reconnection.

The HCS is known as a zone of raised turbulence (Crooker et al. 2004; Blanco et al. 2006). Repeating reconnection produces discontinuities, waves, magnetic islands, and additional local neutral-lines around the main current sheet (Drake et al. 2009; Khabarova & Zastenker 2011). These effects may be responsible for the solar wind turbulence increase with heliocentric distance and give a key for the $B_r$ changing bimodality effect explanation.

Let us show that the magnetic reconnection may take place not only at the HCS but also inside IMF sectors. Indeed, a surprisingly low input of sector boundaries into the IMF components' histograms at 1 AU (Figure 8) is found. The left panel in Figure 8 demonstrates distributions of the three IMF components for 1977–2009 in comparison with the same for the days of sector boundary crossings (SBC). The SBC list by Dr. Leif Svalgaard (http://www.leif.org/research) is taken for calculations. SBC-related distributions are filled with black, and the gray filled curves represent whole time period distributions.

$B_x$, $B_y$, and $B_z$ IMF components are distributed widely on SBC days (Figures 8(a) and (b)), which is a result of a high IMF disturbance level around the HCS. It is remarkable that the distributions have no expected high peaks at zero. To estimate an SBC input into the whole time period histograms, we calculated a relative contribution from the SBC days to each strength span (in percents). Figures 8(d)–(f) show rather significant contribution (up to ~30%) from the HCS to the high values of the $B_x$, $B_y$, and $B_z$ components. At the same time, the SBC contribution to the $B_x$, $B_y$, and $B_z$ distributions for near-zero values is low. It is merely ~15%.

This means that most IMF zeros and weak fields belong to local current sheets that represent former separators between active regions of different sign at the Sun. These thin (in comparison to the HCS) current sheets are located inside sectors, so if the reconnection process occurs not only at the HCS, but everywhere, then it increases sector structure destruction with increasing heliocentric distance.

As a result, farther than 3–4 AU, the turbulent solar wind looks like a sandwich full of randomly distributed neutral lines with alternating by pieces of the inward- or outward-directed IMF. So, under the data average used in this investigation, the observation a probability of a certain IMF strength at a distance from the Sun is approximately determined by the Gaussian function.

Therefore, the magnetic reconnection at the HCS and at local current sheets inside sectors may be another factor that is underestimated, breaking the simple sector structure observed at 1 AU and leading to increasing solar wind complexity with $r$. This effect may be responsible for the fast disappearance of the radial magnetic field component bimodality with distance from the Sun, and for the poor correspondence of the observed IMF expansion law to Parker's model in the inner heliosphere.

## 5. CONCLUSIONS

The observed in-ecliptic IMF strength in the inner heliosphere corresponds poorly to models that use Parker's solution, considering ideal plasma with a "frozen-in" magnetic field that obeys the $r^{-2}$ expansion law for the radial IMF component $B_r$. Based on multi-spacecraft observations from 0.29 AU to 5 AU, we found out that the IMF in the inner heliosphere demonstrates complex behavior depending on heliocentric distance $r$.

1. The IMF radial component $|B_r| \propto r^{-5/3}$, the tangential component $|B_t| \propto r^{-1}$, and the IMF strength $B \propto r^{-1.4}$ with a coefficient of determination very close to 1 ($R^2 = 0.996$). The estimation of the magnetic field power index is very important because it determines the index of the solar wind polytropic equation. According to our calculations, the solar wind polytropic index ~1.38.

The $B_r$ distribution shape varies significantly with the distance from the Sun. Its bimodality is most clearly pronounced at 0.7–2.0 AU, where it represents a mixture of two normal distributions. Meanwhile, the $B_r$ distribution cannot be approximated by two overlapping normal distributions at 0.29–0.4 AU, and it becomes unimodal at 3–4 AU.

The $B_r$ distribution bimodality at 1 AU is a well-known manifestation of the sector pattern. Fast disappearance of the bimodality with increasing distance from the Sun is a sign of increasing solar wind turbulence and structural complexity, which obviously leads to the disappearance of a clear sector structure at 3–4 AU. On the contrary, the determination of the $B_r$ distribution multiple modality at 0.29–0.4 AU possibly means that both IMF sectors and the HCS are too thick at such distances to contribute to the $B_r$ distribution.

Our investigation confirms that the solar wind becomes increasingly turbulent with heliocentric distance, but it was surprising to see that the stable sector structure disappears so fast. The next interesting point is that the IMF radial expansion is determined by the adiabatic power index of −5/3, and not of −2, as is usually supposed. Both the bimodality disappearance effect and the IMF expansion law found to be far from Parker's model may be determined by the breakdown of the "frozen-in IMF" law as well as by a very complex solar wind structure far from the Sun. If our supposition is true, then further consideration of theories using different expansion laws for the solar wind plasma and for the IMF would be preferable for predictive aims.

The magnetic reconnection at current sheets may be responsible for the observed phenomena. Indeed, it was found that the heliospheric current sheet's contribution to the IMF zero values is merely 15%, as observed at the Earth orbit. An overwhelming majority of weak magnetic fields and neutral lines in the solar wind at 1 AU do not belong to the HCS, but rather to local current sheets inside sectors. If magnetic reconnection takes place everywhere in the heliosphere, then it produces current sheet splitting, and causes instabilities and waves to arise not only around the HCS, but also inside sectors, so that the IMF stable sector structure decreases with distance, which is seen through the multi-spacecraft IMF observations.

We have only outlined a part of the whole picture of the IMF spatial and temporal evolution in the inner heliosphere, but our results indicate that the IMF is strongly affected by nonlinear processes in space. The proposed hypotheses are not final and will be checked further. Meanwhile, there is no doubt that the IMF behavior in the inner heliosphere (considered even under a crude approximation) is more complex than often supposed to be.

*OMNI2, Pioneer Venus Orbiter, Helios 2, IMP8, Ulysses, and Voyager 1 data were taken from the official Goddard Space Flight Center OMNIweb plus Web site: http://omniweb.gsfc.nasa.gov. We thank Dr. Leif Svalgaard for the opportunity to use his SBC List (http://www.leif.org/research/sblist.txt). The data of Wilcox Solar Observatory for calculation of the solar wind source surface magnetic field were taken from http://wso.stanford.edu/forms/prsyn.html.*

*This research was supported by the Russian Fund of Basic Researches' grants Nos.11–02-00259-a, 10–02-01063-a, and 12-02-10008-K.*